\documentclass[prb,twocolumn,superscriptaddress,floatfix,showpacs]{revtex4}
\usepackage{graphicx,amsfonts,amssymb,amsmath, hyperref, enumerate}

\newif\ifhyper
\hypertrue
\ifhyper
\hypersetup{
   citecolor = {green},
   colorlinks = {true}, % false
   urlcolor = {blue} % magenta
}
\fi

\newcommand{\beq}{\begin{equation}}
\newcommand{\eeq}{\end{equation}}
\newcommand{\beqa}{\begin{eqnarray}}
\newcommand{\eeqa}{\end{eqnarray}}

\def\Longarrow{\protect\@lra}
\def\@lra{\relbar\joinrel\relbar\joinrel\relbar\joinrel%
          \relbar\joinrel\rightarrow}

\begin{document}

%\title{Nematic versus solid orders in the spin-1/2 Kagome XXZ model in a field}

\title{The spin-1/2 Kagome XXZ model in a field: \\ competition
between lattice nematic and solid orders}

\author{Augustine Kshetrimayum}
\affiliation{Institute of Physics, Johannes Gutenberg University, 55099 Mainz, Germany}

\author{Thibaut Picot}Ê
\affiliation{Laboratoire de Physique Th\'eorique, IRSAMC, CNRS and Universit\'e de Toulouse, UPS, F-31062 Toulouse, France}

\author{Rom\'an Or\'us}
\affiliation{Institute of Physics, Johannes Gutenberg University, 55099 Mainz, Germany}

\author{Didier Poilblanc} 
\affiliation{Laboratoire de Physique Th\'eorique, IRSAMC, CNRS and Universit\'e de Toulouse, UPS, F-31062 Toulouse, France}

\begin{abstract}

We study numerically the spin-1/2 XXZ model in a field on an infinite Kagome lattice. We use different algorithms based on infinite Projected Entangled Pair States (iPEPS) for this, namely: (i) an approach with simplex tensors and 9-site unit cell, and (ii) an approach based on coarse-graining three spins in the Kagome lattice and mapping it to a square-lattice model with local and nearest-neighbor interactions, with usual PEPS tensors, 6- and 12-site unit cells.  Similarly to our previous calculation at the SU(2)-symmetric point (Heisenberg Hamiltonian), for any anisotropy from the  Ising limit to the XY limit, we also observe the emergence of magnetization plateaus as a function of the magnetic field, at  $m_z = \frac{1}{3}$ using 6- 9- and 12-site PEPS unit cells, and at $m_z = \frac{1}{9}, \frac{5}{9}$ and $\frac{7}{9}$ using a 9-site PEPS unit cell, the later set-up being able to accommodate $\sqrt{3} \times \sqrt{3}$ solid order. We also find that,
at $m_z = \frac{1}{3}$, (lattice) nematic and  $\sqrt{3} \times \sqrt{3}$ VBC-order states are degenerate within the accuracy of the 9-site simplex-method, for all anisotropy. The 6- and 12-site coarse-grained PEPS methods 
produce almost-degenerate nematic and $1 \times 2$ VBC-Solid orders. We also find that, within our accuracy, the 6-site coarse-grained PEPS method gives slightly lower energies, which can be explained by the larger amount of entanglement  this approach can handle, even in the cases where the PEPS unit-cell is not commensurate with the expected ground state unit cell. Furthermore, we do not observe chiral spin liquid behaviors at and close to the XY point, as has been recently proposed. Our results are the first tensor network investigations of the XXZ model in a field, and reveal the subtle competition between nearby magnetic orders in numerical simulations of frustrated quantum antiferromagnets, as well as the delicate interplay between energy optimization and symmetry in tensor network numerical simulations.

\end{abstract}

\maketitle

\section{Introduction}

Frustrated quantum antiferromagnets pose some of the hardest numerical challenges in condensed matter physics. This is so because, given frustration, quantum Monte Carlo is hampered by the infamous sign problem. In this context, the Kagome lattice is one of the most famous examples of frustrated lattices given the presence of triangles, where competing antiferromagnetic interactions between nearest-neighbour spins lead to many eigenstates with very different structure but with very similar and low energy, all of them competing to be the true ground state of the system. {ÊThis makes the Kagome lattice an ideal experimental and theoretical setup to have very large quantum fluctuations at low temperatures, specially for spin-1/2 systems with Heisenberg-like interactions, but also for larger spins \cite{fru}.} 

An important model in this context is the the spin-1/2 XXZ model in a field on the Kagome lattice. In such model, an anisotropy angle $\theta \in [0, \frac{\pi}{2} ]$ is introduced that distinguishes the $zz$ interactions between nearest neighbours from the $xx$ and $yy$ ones. For generic values of anisotropy and field the model has $U(1)$ symmetry, and one recovers the full $SU(2)$ symmetry of the Heisenberg model at the isotropic point when the field is absent. The anisotropy allows us to study the crossover from the XY model ($\theta = 0$) to the Ising model ($\theta = \frac{\pi}{2}$). As such, the model exhibits frustration, and in the presence of a magnetic field it gives rise to quantized plateaus in the longitudinal (i.e. along the field) magnetization $m_z$, being the most prominent at $m_z = \frac{1}{3}$, {Êbut also possible at $m_z = \frac{1}{9}, \frac{5}{9}$ and $\frac{7}{9}$ \cite{plateaus}. The existence, nature and properties of such plateaus have been a hot topic of discussion, with some works proposing that in the XY point (i.e. for purely $xx$ and $yy$ interactions) the $m_z = \frac{1}{3}$ plateau may give rise to a chiral topological quantum spin liquid \cite{chirxy}}. In fact, even the existence of the plateau at the XY point is also a matter of controversy. Moreover, it is not clear if the $m_z = \frac{1}{3}$ plateau undergoes a phase transition as one tunes the anisotropy between the XY and the Ising points. 

\begin{figure}
\includegraphics[width=8.6cm,angle=0]{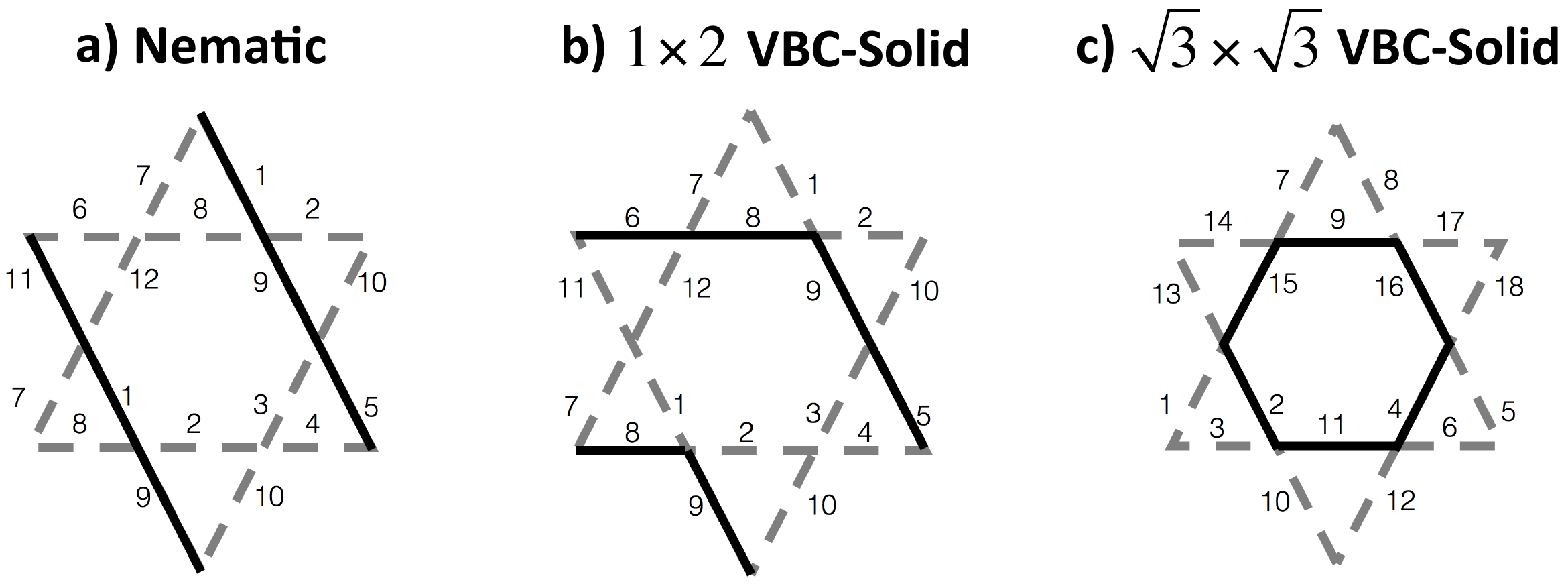}
\caption{[Color online] Different types of order in the Kagome lattice, following the classification from Ref.\cite{heisipeps}: (a) nematic, which breaks $C_6$-symmetry down to $C_2$ and is 3-fold degenerate; (b) $1 \times 2$ VBC-Solid, which breaks (i) $C_6$-symmetry completely, reflection-symmetry about one axis, and (ii) translation-symmetry down to a $1 \times 2$ unit cell; (c) $\sqrt{3}Ê\times \sqrt{3}$ VBC-Solid, which breaks translation-symmetry down to a $\sqrt{3}Ê\times \sqrt{3}$ unit cell and is 3-fold degenerate. { The first two structures in (a) and (b) can be accommodated in a 6-site unit cell, and (c) in a 9-site unit cell. To make this more evident, we show a possible labelling of the different links for such unit cells: 12 links for the 6-site, 18 for the 9-site.}}
\label{fig0}
\end{figure}

In this paper we study the zero-temperature phase diagram of the above model with several methods based on infinite Projected Entangled Pair States (iPEPS) \cite{ipeps}, similarly to previous studies of the Heisenberg point ($\theta = \frac{\pi}{4}$) \cite{heisipeps}. In particular, we use:
\begin{enumerate}[(i)]
\item{An approach with simplex tensors and 9-site unit cell \cite{pess}.}
\item{An approach based on a coarse-graining three spins in the Kagome lattice and mapping it to a square-lattice model with local and nearest-neighbour interactions, with usual PEPS tensors, 6 and 12-site unit cells \cite{heisipeps}.}
\end{enumerate} 

{ As we shall see, the use of different unit cells produces different results in the phase diagrams, depending on whether the magnetic structure of the phase at some plateaus is commensurate with the used cell. Moreover, we shall also see that methods for which the structure of some plateaus is not commensurate may produce, however, slightly lower variational energies because they are able to handle a larger amount of entanglement. In particular, we do a detailed analysis of this situation for the $m_z = 1/3$ plateau. Our different methods reveal a subtle competition between symmetry and energy in the model: while the 9-site unit cell seems better-suited for the expected symmetry properties of the ground state in some parameter regimes such as the $m_z = \frac{1}{9}, \frac{5}{9}$ and $\frac{7}{9}$ plateaus, the approach based on the 6-site unit cell, even if not favourable from the symmetry perspective, turns out to be able to handle more entanglement in the wavefunction and may thus produce slightly better variational energies \footnote{In fact, the property that symmetric tensor networks do not necessarily produce better variational energies is well-known for tensor network methods\cite{pess}.}.} We also observe a competition between different types of order: (i) Valence Bond Crystal solid (VBC-Solid) order breaking lattice translation symmetry and (ii) nematic order 
only breaking $C_3$-rotation symmetry~\cite{heisipeps0}. More specifically, we see that the 9-site simplex-method produces degenerate nematic and $(\sqrt{3} \times \sqrt{3})$ VBC-Solid states \footnote{Notice that in Ref.\cite{heisipeps} this order was called ``Solid 1", in order to distinguish it from other orders with a $\sqrt{3}Ê\times \sqrt{3}$ unit cell at other values of the magnetization.}, within our accuracy, for all values of the anisotropy up to  the Ising point, whereas the 6- and 12-site coarse-grained PEPS methods produce almost degenerate nematic and $(1 \times 2)$ VBC-Solid states, {Êwith a slightly lower (but almost degenerate)} energy than the 9-site approach (see Fig.\ref{fig0} for schematic diagrams of these orders). We also do not find any of the characteristic signatures of a chiral spin liquid phase in the XY point \cite{chirxy}. This paper is also the first tensor network study of the XXZ model in a field on a Kagome lattice. 

\section{Model and methods} Ê

\subsection{The Hamiltonian}Ê
Here we consider the antiferromagnetic XXZ model for spin-1/2 in the presence of an external magnetic field along the $z$ direction on the Kagome lattice. Its Hamiltonian is given by 
\begin{eqnarray}
{\cal H} &=& H - h \sum_i S_i^z, \label{eq1}
\\
H = &\sum_{\langle i j \rangle}&\left( \cos{\theta} \left(S_i^x S_j^x + S_i^y S_j^y \right)  + \sin{\theta}\, S_i^z S_j^z \right),  \nonumber
\end{eqnarray}
where $S_i^\alpha = \sigma_i^\alpha / 2 $ is the spin-1/2 $\alpha$th operator at site $i$, $h$ is the magnetic field, $\theta$ is the anisotropy angle, and the interaction is for nearest-neighbour spins. For generic values of $h$ and $\theta$ this Hamiltonian is $U(1)$-symmetric under spin rotations around the $z$-axis. Different interesting points can be accessed by tuning parameter $\theta$, namely: 
\begin{enumerate}[(i)]
\item{The Heisenberg point for $\theta = \frac{\pi}{4}$.}
\item{The XY point for $\theta = 0$.}
\item{The Ising point for $\theta = \frac{\pi}{2}$.}
\end{enumerate}

{ÊAfter doing an overall analysis of the phase diagram,} we shall focus on the incompressible phase at reduced magnetization $m_z=\frac{1}{3}$ where, in the classical Ising limit, a macroscopic degeneracy occurs between all configurations with two spin up and one spin down (for $h>0$) 
in every triangle \cite{wannier}. When a small XY (ferromagnetic or antiferromagnetic) anisotropy is added,
i.e. $|\theta - \frac{\pi}{2}|\ll 1$, Eq.~(\ref{eq1}) can be mapped to an effective quantum dimer model on the dual hexagonal lattice~\cite{cabra2005} whose ground-state was argued to be a $\sqrt{3} \times \sqrt{3}$ VBC-Solid~\cite{moessner2001}. This result is supported 
by direct Quantum Monte Carlo simulations on the (non-frustrated) hard-core boson model~\cite{isakov2006} equivalent to the ferromagnetic XXZ chain  
given by $\theta>\frac{\pi}{2}$. Here we wish to investigate what happens away from the Ising point { in this case}. 

\subsection{Numerical approaches} 

Following our approach in Ref.\cite{heisipeps}, we use imaginary-time evolution in order to obtain approximations of the ground state of the system in the thermodynamic limit. To implement this, here we use several approaches based on infinite-PEPS (or iPEPS), also similarly to what we did in Ref.\cite{heisipeps}. On the one hand, we use an approach based on introducing simplex tensors, also called ``Projected Entangled Simplex States" (PESS) \cite{pess}, where we use a 9-site unit cell and the so-called ``simple update" \cite{su}.  On the other hand, we also use an approach where three spins in the Kagome lattice are mapped to a single coarse-grained 8-dimensional spin on the square lattice while keeping all interactions local, see Ref.\cite{heisipeps}, and then use a standard PEPS algorithm for the square lattice with 2-site and 4-site square-lattice unit cells, amounting to 6-site or 12-site unit cells in the original Kagome lattice. We have implemented this second approach both with the ``simple update" and also the so-called ``fast full update" \cite{ffu}, but we saw no significant difference in our results with these two update schemes, so we stick to the simple update because it is more efficient. The approach based on the coarse-grained Kagome lattice is less efficient but, however, has  more variational parameters, {Êwhich allows to handle a larger amount of entanglement in the ansatz wavefunction.} For more specific details about these approaches, we refer the reader to Ref.\cite{heisipeps}. 

\begin{figure}
\includegraphics[width=9cm,angle=0]{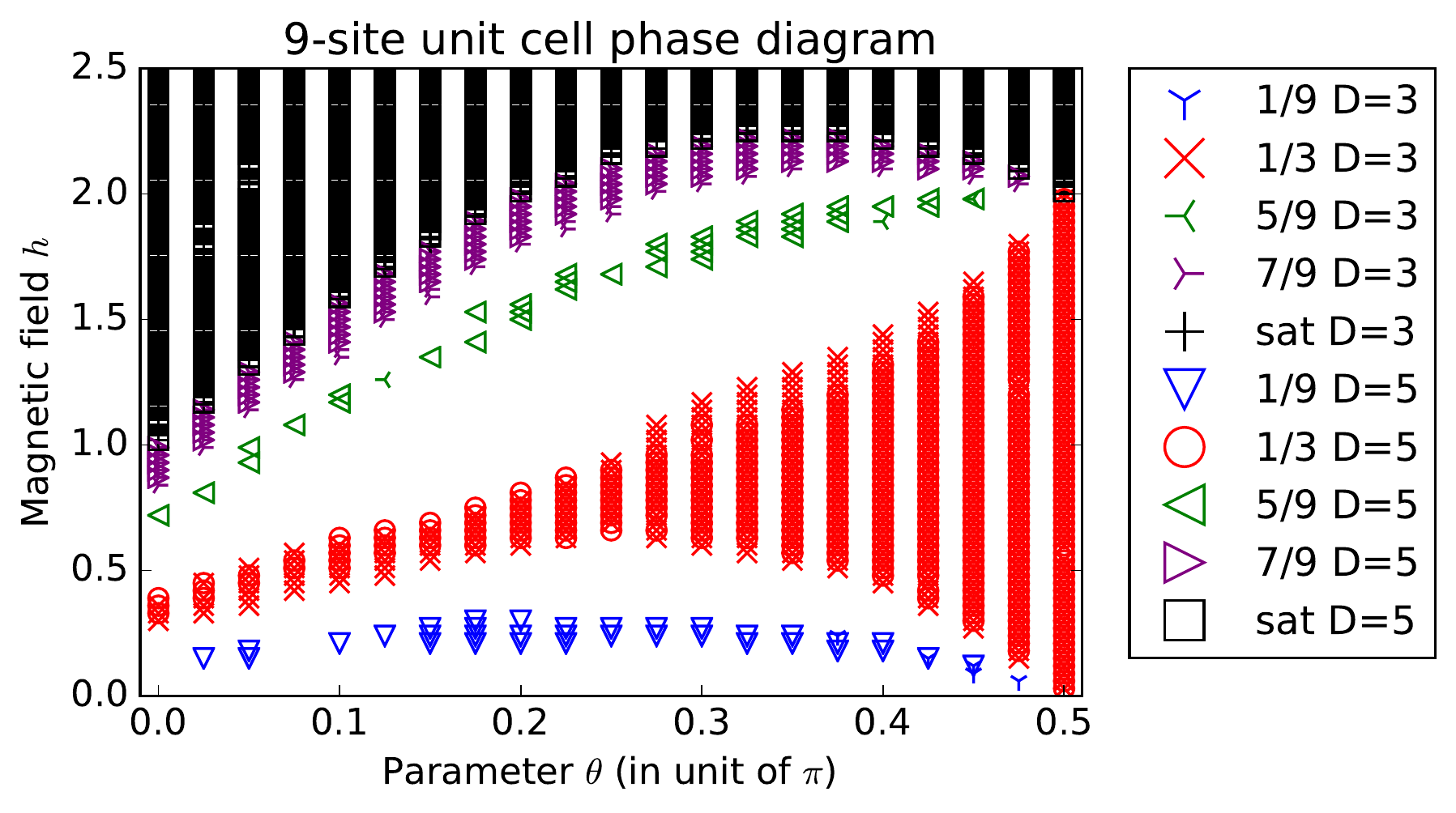}
\caption{[Color online] Different plateaus as a function of the magnetic field $h$ and the anisotropy angle $\theta$, computed with PESS and the 9-site unit cell, for bond dimensions $D=3, 5$.}
\label{fig0_2a}
\end{figure}

In all cases, our refining parameter is the PEPS bond dimension $D$, which controls the amount of entanglement in the tensor network, and which we consider up to $D=5$. Depending on the choice of algorithm, translation symmetry may be broken in different ways, and also the number of variational parameters may be different. For instance, for the PESS approach with a 9-site unit cell, translation symmetry may be broken in, e.g., a $\sqrt{3}\times\sqrt{3}$ superstructure. On the other hand, the coarse-grained lattice approach with 6-site and 12-site unit cells may break translation invariance in other ways, such as a $1 \times 2$ structure. Concerning the number of variational parameters, the coarse-grained approach easily has many more than the PESS approach (e.g. a factor of 4 for $D=3$ and the 6-site unit cell), which means that the calculations need more CPU time, but they {Êare also able to handle more entanglement in the wavefunction}. As in Ref.\cite{heisipeps}, the characterization of the phases is possible by checking the local magnetizations as a function of the field, as well as the link energy terms  $\langle h_{ij}  \rangle$, with $H = \sum_{\langle i j \rangle} h_{ij}$ as in Eq.(\ref{eq1}). Expectation values and environment contractions are computed using Corner Transfer Matrices, see, e.g., Refs.\cite{ctm, tn}. Moreover, in order to check the possible chirality of the wave-function, we compute the entanglement spectra of the state wrapped around half an infinite cylinder, by following the procedure in, e.g, Ref.\cite{chirpeps}.

\begin{figure}
\includegraphics[width=8.8cm,angle=0]{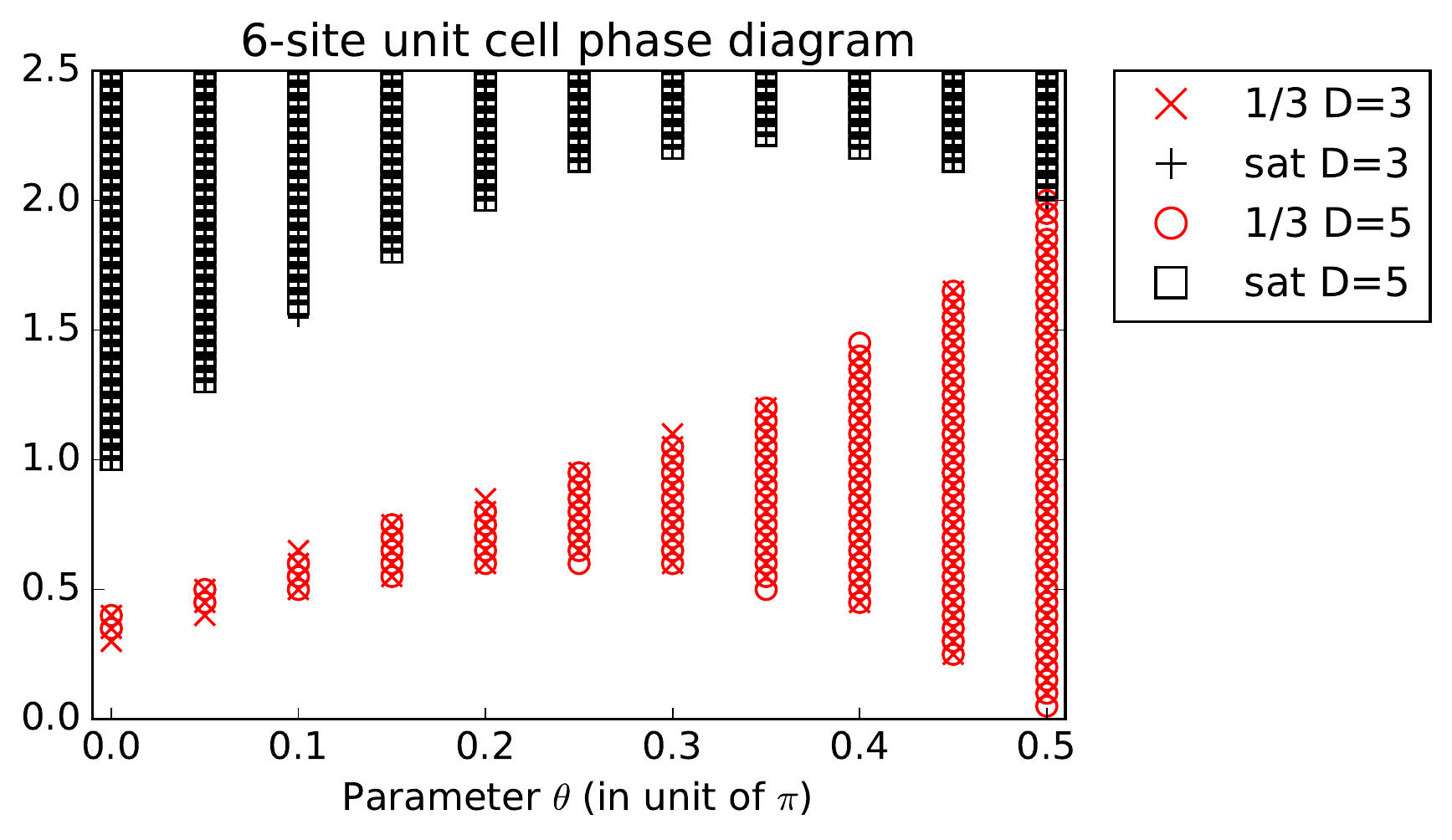}
\caption{[Color online] Different plateaus as a function of the magnetic field $h$ and the anisotropy angle $\theta$, computed with the coarse-grained PEPS and the 6-site unit cell, for bond dimensions $D=3, 5$. The plateaus at $m_z = \frac{1}{9}, \frac{5}{9}$ and $\frac{7}{9}$ appear only for the 9-site cell (see Fig.\ref{fig0_2a}), since the 6-site cell is incommensurate with their magnetic structure.}
\label{fig0_2b}
\end{figure}

\section{Results}Ê

\begin{figure}
\includegraphics[width=8cm,angle=0]{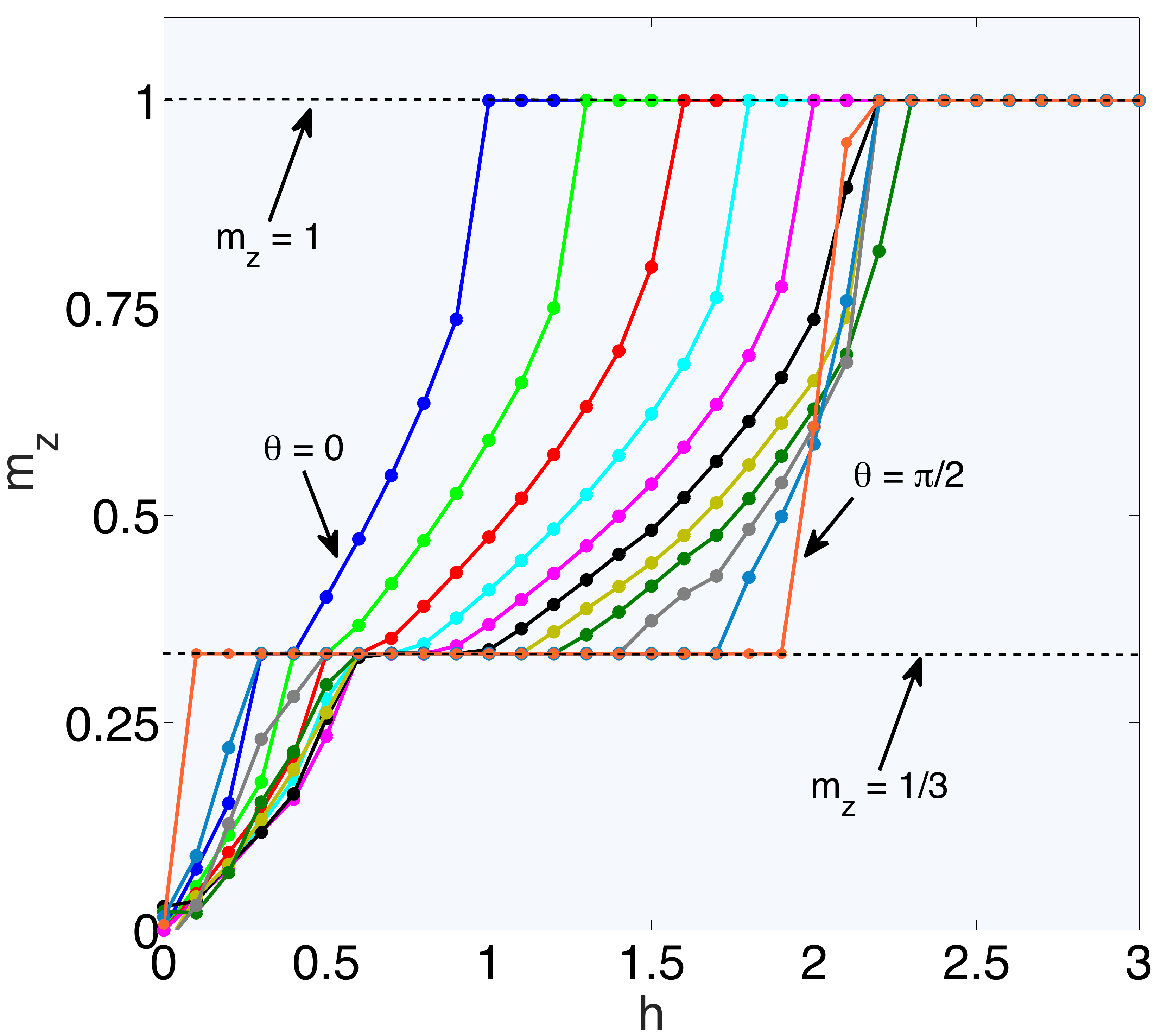}
\caption{[Color online] Magnetization $m_z$ (along the field) as a function of the external magnetic field $h$ in intervals of $\Delta h = 0.1$ for values of the anisotropy angle $\theta \in [0, \pi/2]$ at intervals of $\Delta \theta = \pi/20$, computed with the coarse-graining approach with bond dimension $D=3$ and a 6-site unit cell in the Kagome lattice. Results for higher bond dimensions did not change the picture significantly. The saturation value of the magnetization is $m_z = 1$ (dotted line), and there is clearly a plateau at $m_z = \frac{1}{3}$ (dotted line) for all the considered angles, including the pure XY point at $\theta=0$. {ÊIn this approach the plateaus at $m_z = \frac{1}{9}, \frac{5}{9}$ and $\frac{7}{9}$ are not found since their expected magnetic structure is incommensurable with the unit cell. They are, however, found with the PESS approach and the 9-site unit cell, see Fig.\ref{fig0_2a}.} }
\label{fig1a}
\end{figure}

{ In Fig.\ref{fig0_2a} and Fig.\ref{fig0_2b} we show the different plateaus that arise as a function of the external magnetic field $h$ and the anisotropy angle $\theta$, as computed with the 9-site PESS and the 6-site coarse-grained PEPS. We find that the 9-site PESS reproduces different plateaus at $m_z = \frac{1}{9}, \frac{1}{3}, \frac{5}{9}$ and $\frac{7}{9}$, whereas the 6-site coarse-grained PEPS reproduces only the $m_z = \frac{1}{3}$, due to the incommensurability of this unit cell with the expected magnetic structure of the other plateaus. Concerning the ground state energy, the approach based on the 6-site cell seems to produce slightly lower values than the 9-site cell. For instance, at the $\frac{5}{9}$ plateau region for $\theta = 0.45 \pi$ we find that for $D=3$, the ground state energies per site are  $e_0(6-{\rm site}) \approx -0.520(2)$ (where in fact there is no plateau) and $e_0(9-{\rm site}) \approx -0.519(4)$. This slight energetic advantage may be due to the larger number of variational parameters involved in the 6-site wavefunction for a given $D$. However, variations beyond the third significant digit should be taken with care, since the relative difference is still very small specially in the context of the several approximations involved in both algorithms, and which we control to the best of our possibilities.}  

\begin{figure}
\includegraphics[width=8cm,angle=0]{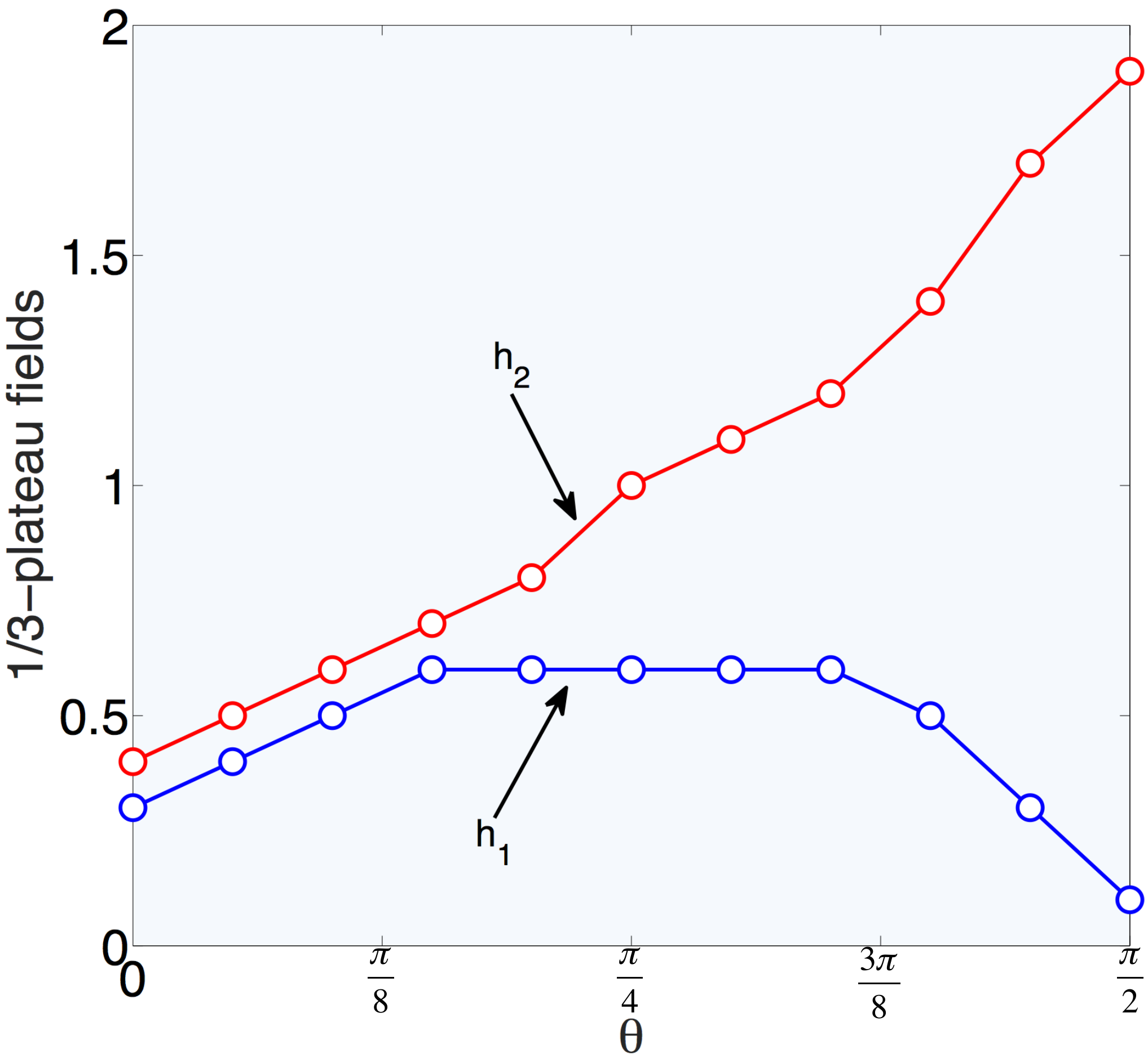}
\caption{[Color online] Left-most ($h_1$) and right-most ($h_2$) values of the field at the $m_z = \frac{1}{3}$ plateau for the different considered angles. The width of the plateau is given by $h_2 - h_1$. The parameters of the simulations are the same as in (a). At the Ising point $\theta = \pi/2$ we find $h_2 - h_1 = 2$ up to our finite-$\Delta h$ resolution ($\Delta h = 0.1$), in agreement with the exact classical result. }
\label{fig1b}
\end{figure}

For concreteness, in Fig.\ref{fig1a} we show the magnetization curve, i.e. $m_z$  as a function of the external field $h$ obtained
with the 6-site algorithm, for different values of the anisotropy angle $\theta \in [0, \pi/2]$. We observe the presence of a very prominent magnetization plateau at $m_z = \frac{1}{3}$, for all the scanned values of $\theta$, including the XY, Heisenberg and Ising points. Away from the Ising point we also see a jump towards the saturation value $m_z = 1$ {Êfor several values of theta \cite{jump} -- a more fine-grained analysis (not displayed) shows that this is indeed the case --.} The width of the plateau increases as we tune $\theta$ from the XY towards the Ising point, see Fig.\ref{fig1b}. Results for the 12-site algorithm are essentially identical. { Notice also that the 9-site PESS algorithm also finds a plateau at the XY point.} \footnote{Note that it was suggested in Ref.\cite{cabra2005} that the plateau width vanishes at $\theta=0$.}

\begin{figure}
\includegraphics[width=8.6cm,angle=0]{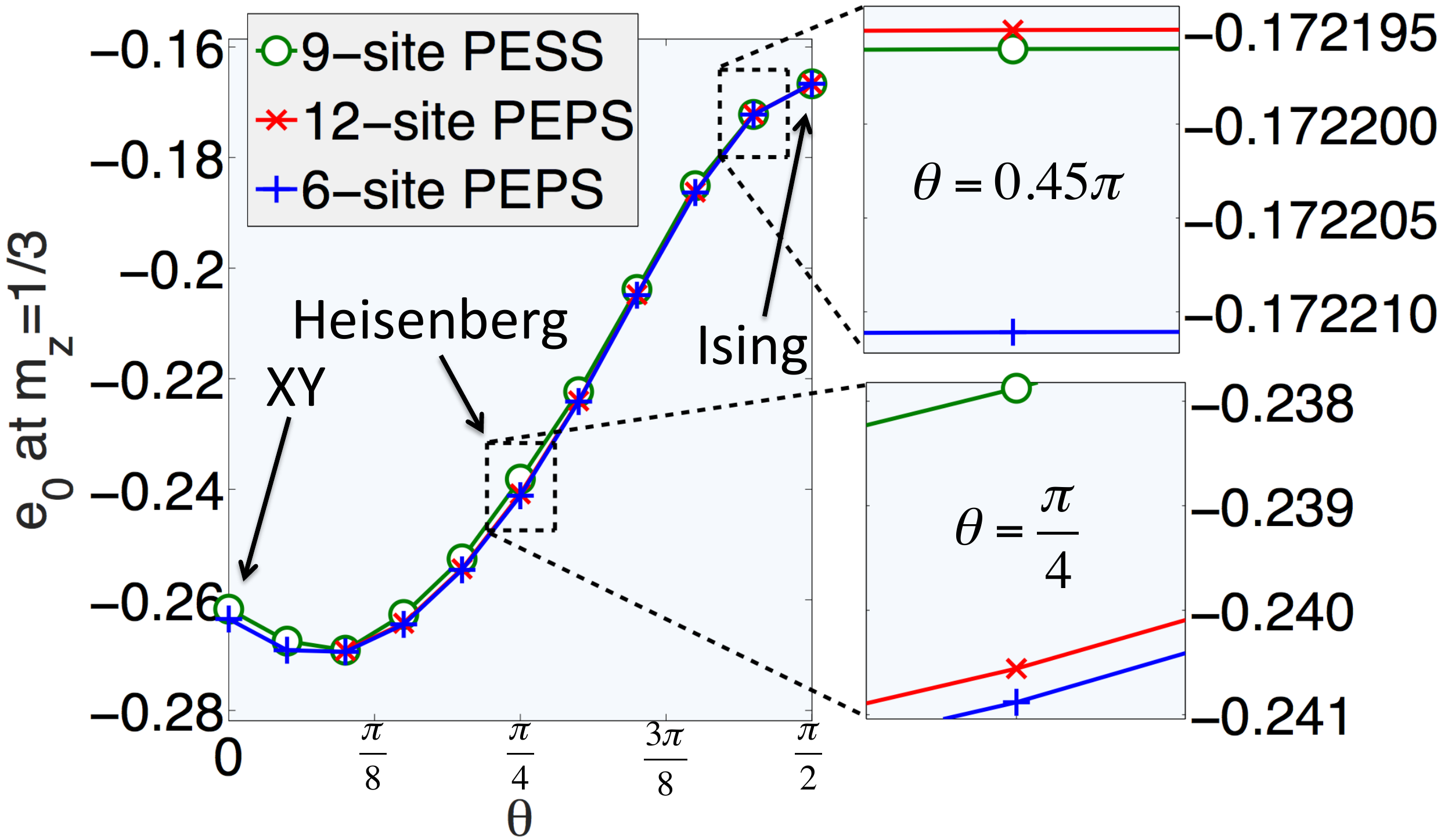}
\caption{[Color online] Comparison of the ground-state energy per site $e_0$ without the Zeeman term at the $m_z = \frac{1}{3}$ plateau for different values of the anisotropy angle $\theta \in [0, \pi/2]$, for different approaches: (i) PESS with 9-site unit cell, (ii) coarse-grained PEPS with 6-site unit cell, and (iii) coarse-grained PEPS with 12-site unit cell. The simulations are for $D=3$ in all cases. The lowest variational energy is given by the coarse-grained PEPS with 6-site unit cell. Higher bond dimensions in the PEPS/PESS lead to the same conclusions when compared on equal footing.ÊThe insets show zooms at the Heisenberg point $\theta = \pi/4$, for which the difference is relatively big, and for $\theta = 0.45 \pi$ close to the Ising point, where the ground state is close to classical (in fact separable at the Ising point $\theta = \pi/2$) and therefore the energy differences are relatively small. }
\label{fig2}
\end{figure}

\begin{figure}
\includegraphics[width=8.6cm,angle=0]{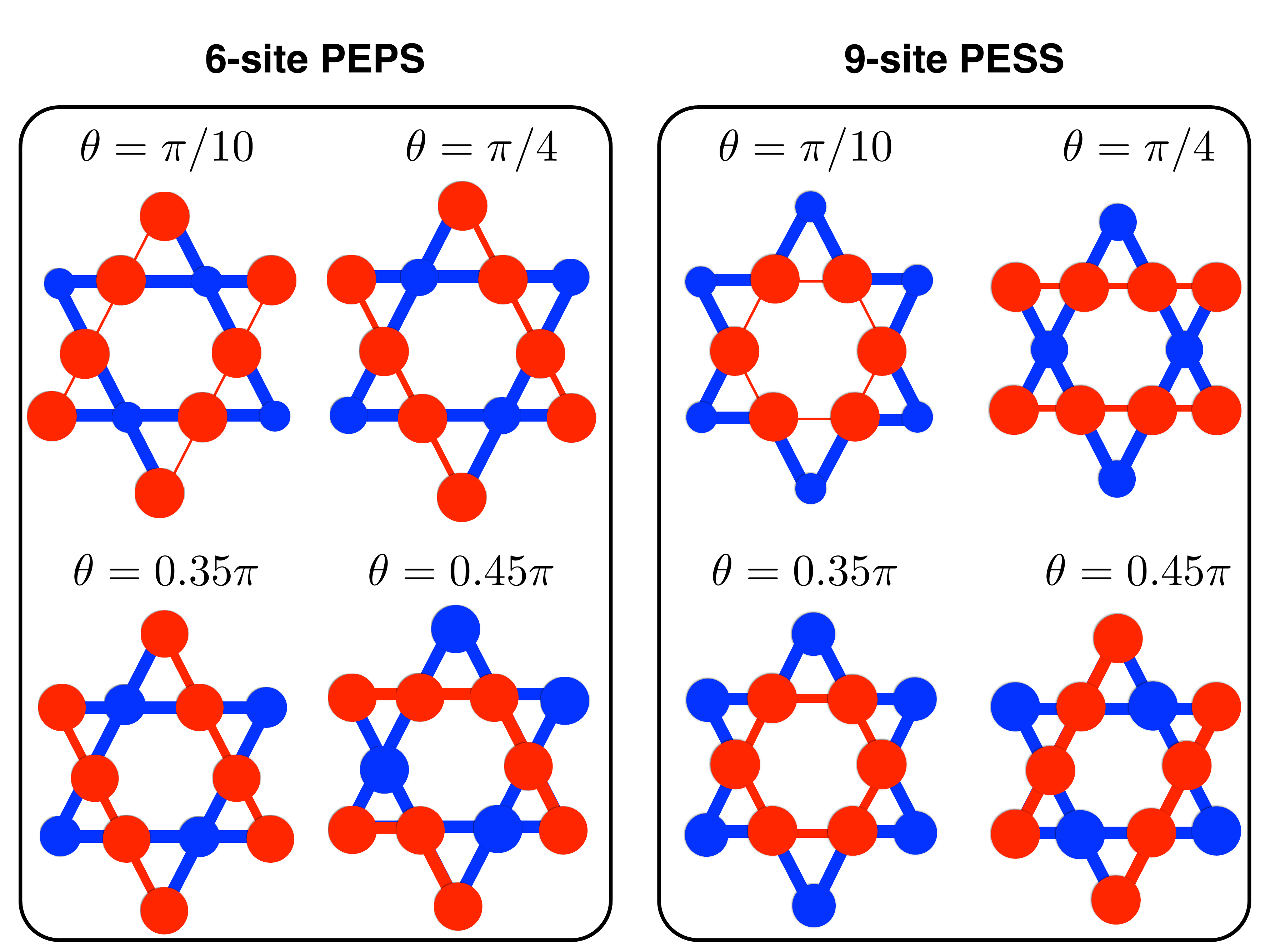}
\caption{[Color online] Magnetizations (along the field) at every site and link energy terms $\langle h_{ij}  \rangle$ at every link, for the coarse-grained approach with 6-site unit cell (6 different sites and 12 different links) (left) and the 9-site PESS approach (9 different sites and 18 different links), and bond dimension $D=3$, for several anisotropy angles. Red means positive, blue means negative, and the thickness indicates the magnitude in absolute value. We observe a clear competition between nematic and VBC-solid orders. For the 6-site PEPS we show nematic for $\theta=\pi/10, \pi/4, 0.35\pi$ and $1 \times 2$ VBC-Solid for $\theta= 0.45\pi$. For the 9-site PESS we see degenerate nematic and $\sqrt{3}\times\sqrt{3}$ VBC-Solid orders for all values of $\theta$ (see text). For illustration, nematic (VBC-Solid) order is shown for $\theta=\pi/4, 0.45\pi$ ($\theta=\pi/10, 0.35\pi$). {ÊThe approximate values (up to three significant digits) of the on-site magnetizations and link energy terms are as follows: (i) for $\theta = \pi/10$: red link $\approx 0.067$, blue link $\approx -0.232$, red dot $\approx 0.653$, blue dot $\approx -0.309$ (same within these digits for both panels);  (ii) for $\theta = \pi/4$: red link $\approx 0.127$, blue link $\approx -0.245$, red dot left $\approx 0.758$, red dot right $\approx 0.736$, blue dot $\approx -0.495$ (same within these digits for both panels expect for the red dot); (iii) for $\theta = 0.35 \pi$: red link $\approx 0.174$, blue link $\approx -0.242$, red dot $\approx 0.858$, blue dot $\approx -0.710$ (same within these digits for both panels); (iv) for $\theta = 0.45 \pi$: red link $\approx 0.237$, blue link $\approx -0.247$, red dot $\approx 0.978$, blue dot $\approx -0.956$ (same within these digits for both panels).}}
\label{fig3}
\end{figure}

\begin{figure}
\includegraphics[width=7.8cm,angle=0]{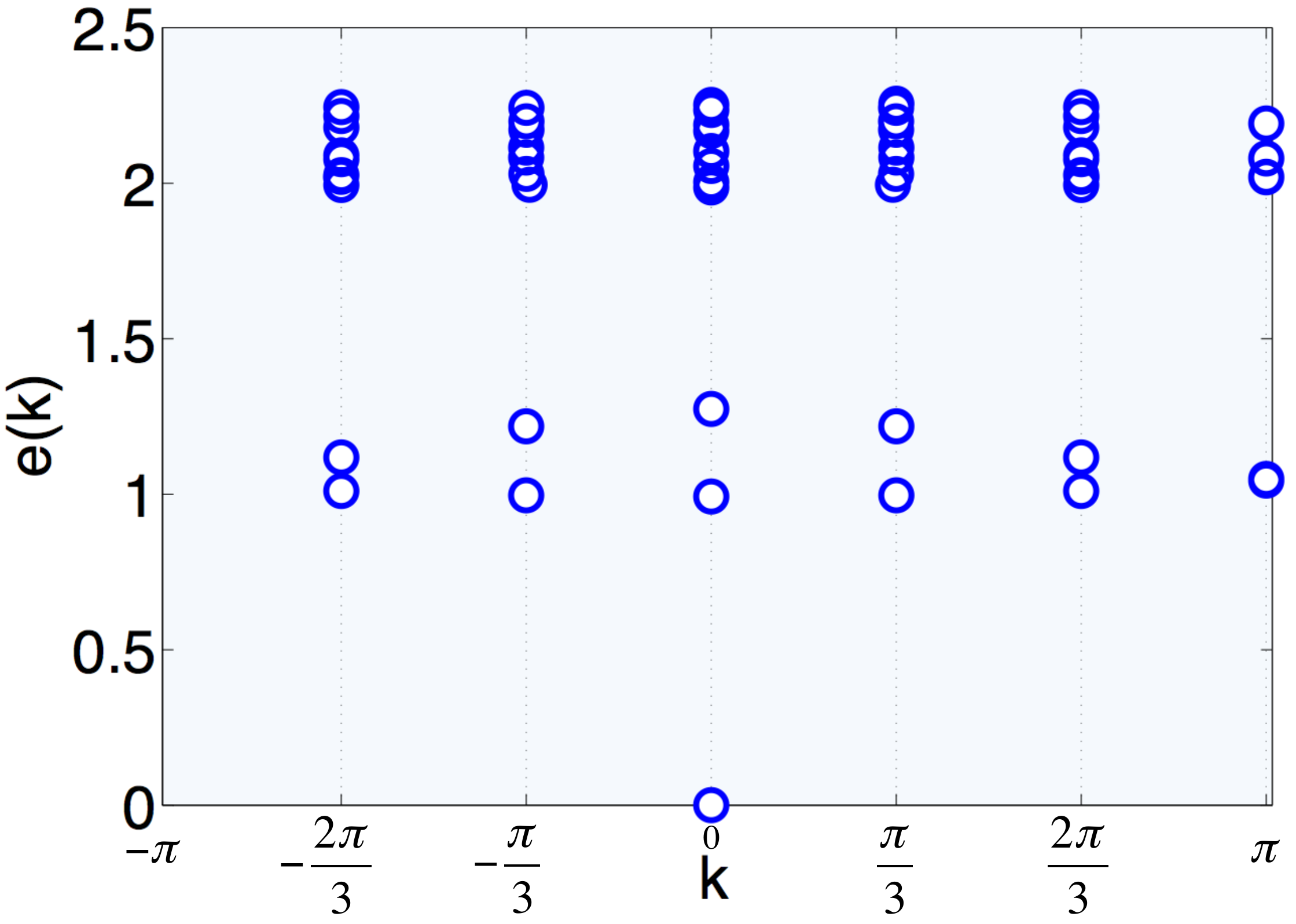}
\caption{[Color online] First 60 ``entanglement energies" of the state at the $m_z = \frac{1}{3}$ plateau for the XY point, obtained by the 6-site PEPS approach with $D=3$. The partition corresponds to half an infinite cylinder with a width of 12 sites, and the cylinder contraction uses a 2-site periodicity as in Ref.\cite{geo}. The spectrum is symmetric in the momentum $k$ in the transverse cylinder direction, and therefore the computed state is non-chiral.}
\label{fig4}
\end{figure}

In Fig.\ref{fig2} we show the ground state energy 
per site of $H$ in the $m_z = \frac{1}{3}$ plateau phase as a function of the anisotropy angle $\theta$, 
 %\emph{without the Zeeman term, which is linear in $h$},
as computed with our three methods. We see that the 6-site approach is the one that produces a lower variational energy. The difference with the 9-site PESS is around the third digit, roughly 1\%, whereas the difference with the 12-site PEPS is even smaller, around 0.1\%. { As said before, these small differences should be taken with caution, given the different approximations involved in the methods used.} The fact that the 9-site PESS approach produces higher energy may be a consequence of the smaller number of variational parameters in this ansatz. The quasi-degeneracy between the 6-site and the 12-site PEPS energies shows that  a doubling of the unit cell is preferred over a quadrupling.  We have also observed that the 9-site PESS produces degenerate nematic and VBC-Solid states (within our energy resolution), similar to what we found in Ref.\cite{heisipeps} at the Heisenberg point. This degeneracy is very slightly lifted in the 6- and 12-site PEPS approaches, in favor of the nematic order for XY anisotropy, in the vicinity of the Heisenberg point
(with a typically relative energy difference of $\sim 0.1\%$ at the Heisenberg point) and for a small Ising anisotropy, and in favor of VBC-Solid order close to the Ising point. { This is true even for the largest bond dimensions that we could reach. It may be possible that the energy difference between these states becomes narrower in the large-size limit, thus explaining this observation.} The stability of one phase w.r.t. the other is assessed when systematically obtaining the same one in different runs of our algorithms with different initial conditions. 

We next perform an analysis of the $m_z = \frac{1}{3}$ plateau {Êwith the different algorithms}. In Fig.\ref{fig3} we show diagrammatically the longitudinal magnetizations at every site as well as the link energy terms $\langle h_{ij}  \rangle$ at every different link, for the 6-site PEPS (left) and the 9-site PESS (right) calculations, for points in the $m_z = \frac{1}{3}$ plateau and four different anisotropy angles $\theta$. In our simulations we have seen that, for the 6-site PEPS approach, for $\theta \lesssim 0.45 \pi$ the phase corresponds, according to the classification in Ref.\cite{heisipeps}, to a nematic state, where $C_6$ rotation symmetry is broken down to $C_2$ (3-fold degenerate). For $\theta \gtrsim 0.45 \pi$ we obtain a VBC-Solid phase (again according to the classification in Ref.\cite{heisipeps}), where translation symmetry is broken down to a $1 \times 2$ cell, and the discrete symmetries $C_6$ (lattice 6-fold rotation) and $\sigma_v$ (reflexion about one axis) are fully broken. Note that the Heisenberg point $\theta = \pi/4$ lies then within the nematic region. Comparison with the results obtained from the 9-site PESS approach is useful. In the right panel of Fig.~\ref{fig3} we show different orders obtained in this way.  In practice we have seen that, for every possible $\theta$, nematic and VBC-solid orders are degenerate in energy within our accuracy. The structures in the figure are examples of what we obtain in some runs of the algorithm. By running the algorithm with different random initial conditions, we get either a nematic or $\sqrt{3}\times\sqrt{3}$ VBC-Solid state, both with the same energy. Such a degeneracy between nematic and VBC-Solid seems to be lifted slightly in the 6- and 12-site PEPS approaches. 

Finally, in order to see possible signs of chirality, we have computed the entanglement spectrum of the PEPS obtained from the 6-site PESS, for half an infinite cylinder of width 12, within the $m_z = \frac{1}{3}$ plateau at the XY point. Such state has been conjectured to be chiral \cite{chirxy}. In our simulation, however, the entanglement spectrum is perfectly symmetric under time-reversal and therefore not chiral, see Fig.\ref{fig4} \footnote{{ÊAs a word of caution, it is well-known that finding a gapped chiral topological state with PEPS may actually be quite difficult because of fundamental constraints of the PEPS tensor network, and therefore it is also difficult to get an intuition about from a numerical optimization.}}

\section{Conclusions}
Ê
Here we have studied the zero-temperature phase diagram of the Kagome XXZ model in a field and in the thermodynamic limit using different tensor network approaches. { We find different plateau structures for the longitudinal magnetization as a function of the field and the anisotropy, depending on the unit cell being used. We also find that a 6-site coarse-grained PEPS approach seems to produce slightly lower ground state energies, probably because the larger number of available parameters which allows for more entanglement in the ansatz wavefunction, and in spite of the fact that the unit cell cannot accommodate the $\sqrt{3}\times\sqrt{3}$ superstructures found in other
recent work~\cite{huerga2016}. Such small energy differences should be considered with caution, though, given the different approximation schemes involved in each method.} From our results, we cannot discern if our 6-site 
ansatz, which contains more adjustable parameters, slightly bias the results towards 
a 6-site superstructure, or if this superstructure is indeed a genuine feature of the system.
In any case, our results show a tight competition between nematic and VBC-Solid orders, which we find for all values of the anisotropy, {Êas well as the delicate interplay between the implementation of (lattice) symmetries and the optimization of the energy in tensor network algorithms.}

\acknowledgements
A. K. and R. O. acknowledge financial support from JGU, as well as the University of Toulouse and the CCBPP, where part of this work was performed. Discussions with Y. Iqbal and M. Ziegler are also acknowledged.

\end{document}